\documentclass[final,5p,times,twocolumn]{elsarticle}

\usepackage{hyperref,longtable}

\journal{Journal of \LaTeX\ Templates}
\biboptions{square,sort,comma,numbers}

\begin{document}

\begin{frontmatter}

\title{The \textsc{Majorana Demonstrator} calibration system}

\author[lbnl]{N.~Abgrall}
\author[pnnl]{I.J.~Arnquist}
\author[usc,ornl]{F.T.~Avignone~III}
\author[ITEP]{A.S.~Barabash}
\author[ornl]{F.E.~Bertrand}
\author[lanl]{M.~Boswell}
\author[lbnl]{A.W.~Bradley}
\author[JINR]{V.~Brudanin}
\author[duke,tunl]{M.~Busch}
\author[uw]{M.~Buuck}
\author[unc,tunl]{T.S.~Caldwell}
\author[sdsmt]{C.D.~Christofferson}
\author[lanl]{P.-H.~Chu}
\author[uw]{C. Cuesta}
\author[uw]{J.A.~Detwiler}
\author[sdsmt]{C.~Dunagan}
\author[ut]{Yu.~Efremenko}
\author[ou]{H.~Ejiri}
\author[lanl]{S.R.~Elliott}
\author[uw]{Z.~Fu}
\author[lanl]{V.M.~Gehman}
\author[unc,tunl]{T.~Gilliss}
\author[princeton]{G.K.~Giovanetti}
\author[lanl]{J.~Goett}
\author[ncsu,tunl,ornl]{M.P.~Green}
\author[uw]{J.~Gruszko}
\author[uw]{I.S.~Guinn}
\author[usc]{V.E.~Guiseppe}
\author[unc,tunl]{C.R.~Haufe}
\author[unc,tunl]{R.~Henning}
\author[pnnl]{E.W.~Hoppe}
\author[unc,tunl]{M.A.~Howe}
\author[usd]{B.R.~Jasinski}
\author[blhill]{K.J.~Keeter}
\author[ttu]{M.F.~Kidd}
\author[ITEP]{S.I.~Konovalov}
\author[pnnl]{R.T.~Kouzes}
\author[ut]{A.M.~Lopez}
\author[unc,tunl]{J.~MacMullin}
\author[queens]{R.D.~Martin}
\author[lanl]{R. Massarczyk\corref{ca}}
\cortext[ca]{Corresponding author}
\ead{massarczyk@lanl.gov}
\author[unc,tunl]{S.J.~Meijer}
\author[lbnl]{S.~Mertens}
\author[pnnl]{J.L.~Orrell}
\author[unc,tunl]{C.~O'Shaughnessy}
\author[lbnl]{A.W.P.~Poon}
\author[ornl]{D.C.~Radford}
\author[unc,tunl]{J.~Rager}
\author[unc,tunl]{A.L.~Reine}
\author[lanl]{K.~Rielage}
\author[uw]{R.G.H.~Robertson}
\author[unc,tunl]{B.~Shanks}
\author[JINR]{M.~Shirchenko}
\author[sdsmt]{A.M.~Suriano}
\author[usc]{D.~Tedeschi}
\author[unc,tunl]{J.E.~Trimble}
\author[ornl]{R.L.~Varner}
\author[JINR]{S.~Vasilyev}
\author[lbnl]{K.~Vetter\fnref{ucb}}
\author[unc,tunl]{K.~Vorren}
\author[lanl]{B.R.~White}
\author[unc,tunl,ornl]{J.F.~Wilkerson}
\author[usc]{C.~Wiseman}
\author[usd]{W.~Xu}
\author[ornl]{C.-H.~Yu}
\author[ITEP]{V.~Yumatov}
\author[JINR]{I.~Zhitnikov}
\author[lanl]{B.X.~Zhu}

\address[lbnl]{Nuclear Science Division, Lawrence Berkeley National Laboratory, Berkeley, CA, USA} 
\address[pnnl]{Pacific Northwest National Laboratory, Richland, WA, USA}
\address[usc]{Department of Physics and Astronomy, University of South Carolina, Columbia, SC, USA}
\address[ornl]{Oak Ridge National Laboratory, Oak Ridge, TN, USA} 
\address[ITEP]{National Research Center ``Kurchatov Institute'' Institute for Theoretical and Experimental Physics,
Moscow, Russia} 
\address[lanl]{Los Alamos National Laboratory, Los Alamos, NM, USA} 
\address[JINR]{Joint Institute for Nuclear Research, Dubna, Russia}
\address[duke]{Department of Physics, Duke University, Durham, NC, USA}
\address[tunl]{Triangle Universities Nuclear Laboratory, Durham, NC, USA}
\address[uw]{Center for Experimental Nuclear Physics and Astrophysics, and Department of Physics, University of
Washington, Seattle, WA, USA}
\address[unc]{Department of Physics and Astronomy, University of North Carolina, Chapel Hill, NC, USA}
\address[sdsmt]{South Dakota School of Mines and Technology, Rapid City, SD, USA}
\address[ut]{Department of Physics and Astronomy, University of Tennessee, Knoxville, TN, USA} 
\address[ou]{Research Center for Nuclear Physics and Department of Physics, Osaka University, Ibaraki, Osaka, Japan}
\address[ncsu]{Department of Physics, North Carolina State University, Raleigh, NC, USA} 
\address[princeton]{Department of Physics, Princeton University, Princeton, NJ, USA}
\address[usd]{Department of Physics, University of South Dakota, Vermillion, SD, USA} 
\address[blhill]{Department of Physics, Black Hills State University, Spearfish, SD, USA}
\address[ttu]{Tennessee Tech University, Cookeville, TN, USA} 
\address[queens]{Department of Physics, Engineering Physics and Astronomy, Queen's University, Kingston, ON, Canada} 
\fntext[ucb]{Alternate Address: Department of Nuclear Engineering, University of California, Berkeley, CA, USA}

\begin{abstract}
The \textsc{Majorana} Collaboration is searching for the neutrinoless double-beta decay of the nucleus
$^{76}$Ge. The \textsc{Majorana Demonstrator} is an array of germanium detectors deployed with the aim of implementing
background reduction techniques suitable for a 1-tonne $^{76}$Ge-based search. The ultra low-background
conditions require regular calibrations to verify proper function of the detectors.
Radioactive line sources can be deployed around the cryostats containing the detectors for regular energy calibrations.
When measuring in low-background mode, these line sources have to be stored outside the shielding so they do not
contribute to the background. The deployment and the retraction of the source are designed to be controlled by the data
acquisition system and do not require any direct human interaction. In this paper, we detail the design requirements and
implementation of the calibration apparatus, which provides the event rates needed to define the pulse-shape cuts and
energy calibration used in the final analysis as well as data that can be compared to simulations.
\end{abstract}


\end{frontmatter}

\section{Introduction}
Neutrinoless double-beta decay ($0\nu\beta\beta$) is a hypothesized yet unobserved second order process not permitted
by the Standard Model. Such a second order weak process would violate lepton number
conservation\,\cite{Avignone2008,Vergados2012}.
Completed searches to date and first results from running experiments
\cite{Klapdor2006,Exo2014,Agostini2016,Gando2017} indicate
half lives extending beyond 10$^{25}$\,years.
If this process were to be observed experimentally, it would signal the Majorana nature of the neutrinos and indicate a
violation of the lepton number. Through sphaleron processes in the early universe, the violation of baryon number would
then result, a necessary condition and explanation for the present-day matter excess over antimatter in the universe
\cite{Cohen1993}.\\
The \textsc{Majorana Demonstrator} (MJD)\,\cite{Abgrall2014} is a research and development effort aimed at deploying
novel background reduction techniques. Such techniques are needed to build a 1-ton experiment with a projected
background rate after analysis cuts of less than 1\,count/(ROI· tonne · year) at the
Q-value of the 0$\nu\beta\beta$ decay at 2039\,keV.
This is accomplished by fielding an array of highly enriched p-type point contact (PPC) Ge detectors underground at the
4850\,ft level of the Sanford Underground Research Facility with special attention to the radio-purity of
materials in the environment surrounding the Ge detectors. The array is divided into two cryostats, each containing 7
strings of 4 or 5 detectors. Each of the two cryostats has its own vacuum and cooling system. These independent
assemblies are referred to as Modules 1 and 2. Twenty nine kg of the detectors are made of enriched Ge material that is
enriched to $>$87$\%$ in $^{76}$Ge and 15 kg from natural Ge.
In this paper, we describe the design requirements and implementation of a system for calibrating the detectors in
energy and providing high statistics samples for pulse-shape analysis.

\section{Requirements and design}
The calibration system must provide events from a known radioactive source to each detector in the array in about
an hour long data set. This ensures that regular calibrations do not significantly reduce the amount of
live time needed for the physics program of the \textsc{Demonstrator}. The current MJD commissioning and data-taking
plan foresees one to two weekly calibrations. The dynamic range of the germanium detector and the attached read-out
chain goes from sub keV up to 3 MeV in high gain channels and from about 10 keV to 10 MeV in low gain channels. Our
region of interest (ROI) is a 3-keV window at 2039 keV, which is the Q-value of a neutrinoless double-beta decay for
$^{76}$Ge. Therefore, a $^{228}$Th source is suitable for energy calibration using regions below and above the ROI.
Furthermore, it provides peaks for energy calibration at lower energies, which is important for the analysis and the
understanding of the background spectrum.\\
Such a source produces gamma rays with a variety of energies. The intensity and relative location of the full-energy
peaks, see Table\,\ref{tab:energies}, are used to provide a reliable energy calibration for low energies using Pb x-rays
around 80\,keV as well as at higher energies using the $^{208}$Tl line at 2614\,keV. For this highest energy, we aim for
the uncertainty in the number of events in the full-energy peak to be less then 1\%, so that fits of the peak shape
allow an accurate determination of the energy. Adequate statistics during one calibration are determined by the number
of events in the $^{228}$Th decay chain peaks. Its 1.9-yr half-life is sufficient to provide enough statistics even
through the end of MJD operations to periodically evaluate pulse-shape cut efficiencies and detector stability.\\
\begin{table}[h!]
\centering
 \begin{tabular}{c c c}
  \hline
 Energy (keV) & isotope & intensity per decay\\
 \hline
 \hline
238.63 & $^{212}$Pb & 0.433 \\
240.99 & $^{224}$Ra & 0.041 \\
277.36 & $^{208}$Tl & 0.023 \\
300.09 & $^{212}$Pb & 0.032 \\
583.19 & $^{208}$Tl & 0.304 \\
727.33 & $^{212}$Bi & 0.065 \\
785.37 & $^{212}$Bi & 0.011 \\
860.56 & $^{208}$Tl & 0.044 \\
2614.53 & $^{208}$Tl & 0.356 \\
\hline
\hline
\end{tabular}
\caption{Overview on most important gamma energies used for MJD calibration. The single and double-escape peak of the
2614-keV line are used as a test for single-site and multi-site events }
\label{tab:energies}
\end{table}
\newline

\begin{figure}[t]
 \centering
 \includegraphics[width=1.0\columnwidth,bb=0 0 632
472,keepaspectratio=true]{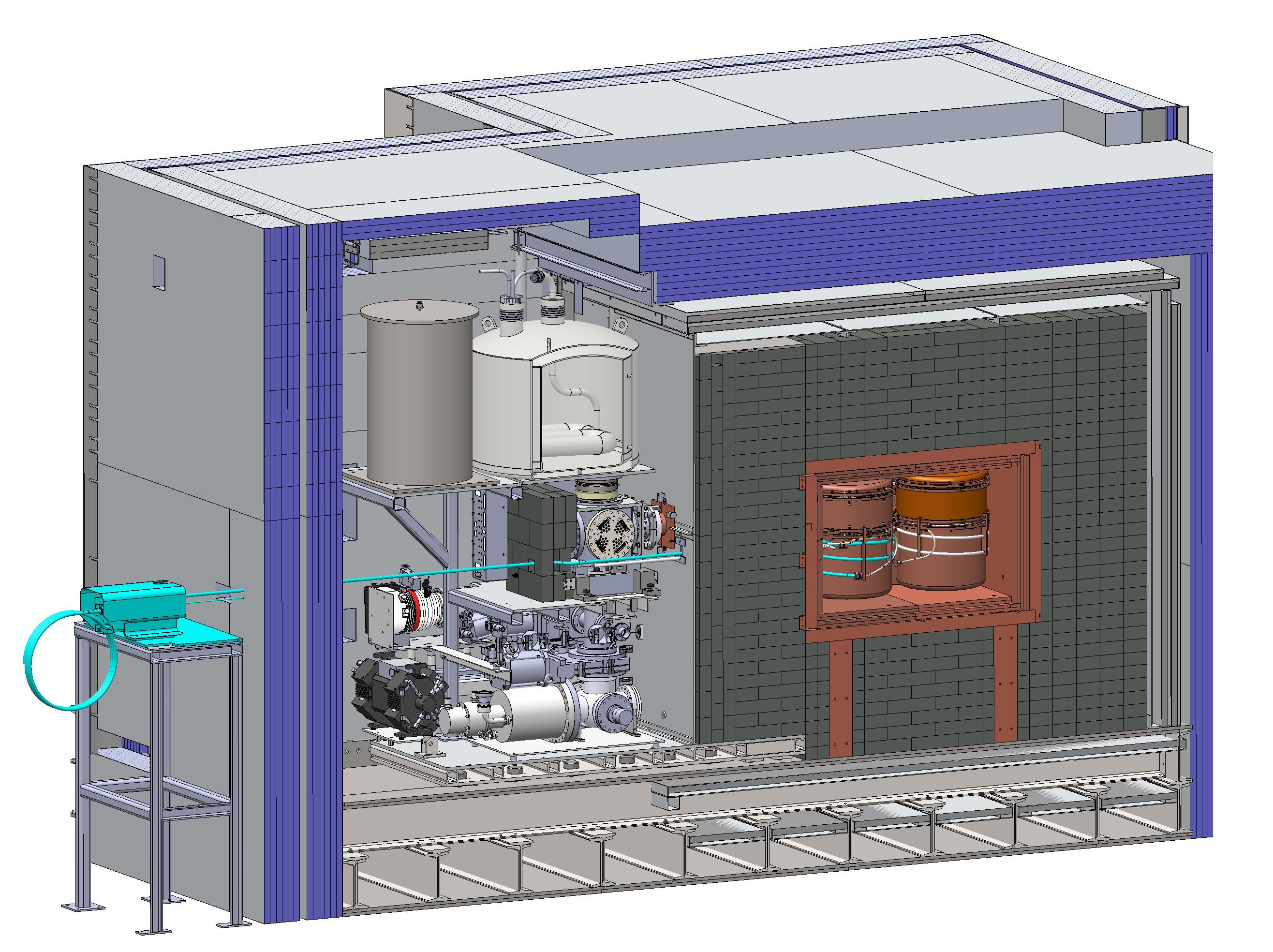}
 \caption{Drawing of a \textsc{Majorana} module assembly with calibration system at final position, including the
support structure, the vacuum system, nitrogen supply as well as the shielding layers. The loops
of the track of Module 2 calibration are visible in cyan. The calibration controls are next to it under the cover next
to the loops of the outer storage track. }
 \label{Fig_1}
\end{figure}

\begin{figure}[t]
 \centering

 \includegraphics[width=1.0\columnwidth,bb=0 0 2200 1700,keepaspectratio=true]{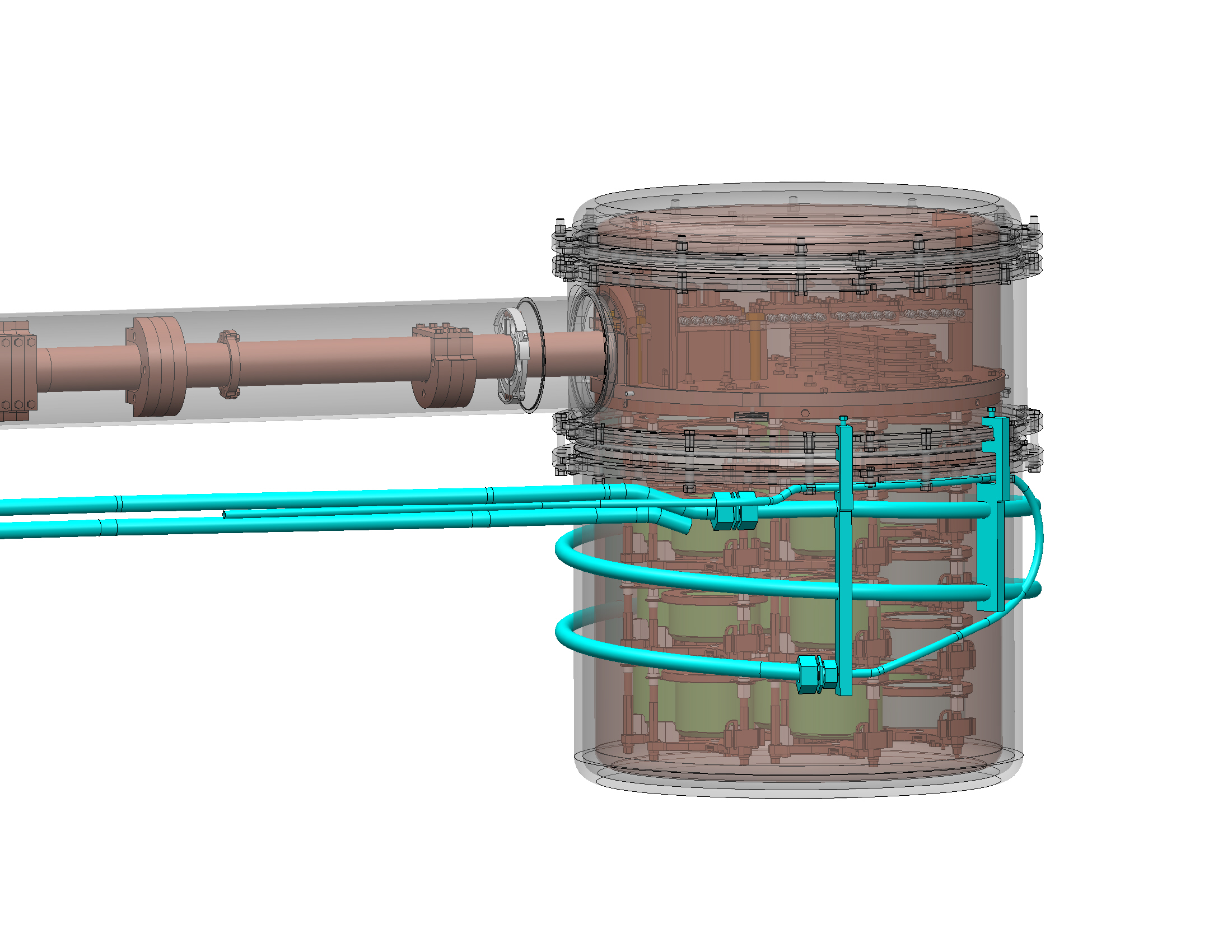}
 \caption{Close-up of the drawing in Fig.\ref{Fig_1} which shows the relative location of detector array and
calibration module.}
 \label{Fig_2}
\end{figure}

Since multiple energy depositions in a single detector
(multi-site events) can sum to energies in our region of interest, we are making use of the unique pulse
shape discrimination abilities of PPC detectors to identify single and multi-site events\,\cite{Cooper2011,
Mertens2015}. Hence additional requirement is the provision
of enough data to test the efficiency of pulse-shape discrimination cuts in the final analysis. The gain of Ge detectors
is generally stable, and can be monitored closely via use of an external pulser.
However, a regular calibration provides an additional cross-check of the long-term stability of the setup. Detector
resolution as well as linearity are monitored by calibration runs. \\
In addition to basic physics requirements, we are technically limited to a total event rate that can be
handled by the front end electronics without excessive pileup. The characteristic slow charge collection
of point contact detectors, which makes pulse-shape discrimination possible, in conjunction with the resistive
feedback of our front end, limits the allowed event rate to roughly 100 Hz per detector. Detectors on the
outermost circumference of the array lie closer to any source deployed around the vacuum cryostat and see
a higher rate than those in the center. Therefore, a compromise must be struck between source
activity, the position and extent of the source, and the time allotted for calibration runs.\\
The modular design of the MJD cryostats in combination with the general construction approach have
led us to adopt a line source that is deployed through the \textsc{Demonstrator}'s Pb and Cu shielding in a helical
track that surrounds each cryostat module, see Figs.\ref{Fig_1}, \ref{Fig_2} and \ref{Fig_3}. In
such a design, the radioactive material is distributed along a cylindrical line-shaped container which can be moved
along a track.\\
The results of a Geant4\,\cite{Agostinelli2003} simulation were used to determine the optimum source
geometry and activity. The simulations
considered different choices of pitch angle for a source along a helical path around the cryostat.
Results indicated that shielding and symmetry considerations produce roughly equivalent rates in the
external detectors, and the driving consideration becomes the event rate in the central strings. Though the
solid-angle exposure to the source is consistent for all detectors of an array, there is a greater
shielding effect to the top and bottom center detectors, see Fig.\,\ref{Fig_2}. Variations in detector dimension across
the range of our available detector sizes had a negligible effect on event rate. For this analysis the
double-escape peak of the 2614-keV line in $^{208}$Tl is used which is part of the decay chain of  $^{228}$Th. From the
simulations \cite{Boswell2010b} we
conclude that calibration runs should contain at least 400 events in this double-escape peak (as a proxy for
the single site analogue for $0\nu\beta\beta$) to validate the pulse-shape cut efficiency.
To understand the behaviour and the relative number of single-escape gamma lines in the Compton shoulder, a $^{60}$Co
source is used in the commissioning phase. With the maximum rate in the outer detectors a thorium source of
$\sim$10\,kBq is
needed so that the sufficient number of events can be reached in calibration data set of roughly one hour in length.

\section{Mechanical implementation}


\subsection{Radioactive line sources}
Each line source consists of a radioactively-doped epoxy injected into a 3-mm-diameter tube that is sealed at
both ends, produced to our custom specifications by Eckert $\&$ Ziegler Analytics,
Inc\footnote{http://www.ezag.com/home/}. The source activities
are conveniently tuned at production; once cured, can be treated as a sealed source.
Deviations from homogeneity along the length of the source are below 3\%.
Four $^{228}$Th calibration sources were prepared in 1-meter lengths, each with an integrated activity of 5.18$\pm$0.30
kBq. A single $^{60}$Co source for building the PSA library was prepared with an integrated activity of 6.3$\pm$0.30 kBq
over a 2 meter length. The $^{60}$Co source can be exchanged between the two \textsc{MJD} modules at their respective
commissioning times. In the final configuration, we fabricated 3 line-source assemblies: one that consists of only
the $^{60}$Co source and two assemblies each consisting of a pair of  $^{228}$Th sources. For each module only one
source at a time can be used.\\
For positioning purposes, a series of 5-mm long NdFeB grade N42 magnetic slugs are embedded in the source tubing, one
pair at the front end of the source, and a second pair with a different spacing between the magnets at the trailing end.
The positions of these magnets could be sensed by Hall-effect devices located outside the tubing. Inside the line-source
assembly the position of the source containers and magnets are fixed using epoxy. The wall thickness of the outer Teflon
containment which encapsulates the source material and the magnets is 0.25 mm. The container has been leached in
agreement with the MJD cleanliness procedures\,\cite{Abgrall2016}. During the cleaning steps, a leach removes surface
contamination and ensures in case of abrasions of the tube material, no radioactive contamination remains close to the
detectors. The total as-built length of a line source assembly is about 4.7 meters.\\

\subsection{Cleanliness}
The MJD shield consists of several layers of different shielding material. Electro-formed copper, commercial copper,
lead and poly complete the shielding and protect the detectors against radiation of natural origin. A radon
exclusion box covers the inner three layers and is purged at all times with boiled-off nitrogen. Around this, two layers
of plastic scintillator are used as an active muon veto. The calibration source has to pass through all of these layers;
its support structure is located outside the shielding behind the poly shield, see
Fig.\ref{Fig_1}. The line source assembly moves in a purged track around the cryostat. This track is made of
polytetrafluoroethylene (PTFE). It was assayed in the MJD assay program and leached prior to installation. The chosen
PTFE\footnote{http://www.coleparmer.com} has a very small natural radioactivity with mass fractions below 3.1$\cdot$
10$^{-12}$ for $^{238}$U and around 1.5$\cdot$10$^{-12}$ for $^{232}$Th.
Inside the shielding, a nitrogen purge line is attached at the inner end to the track and is actively purging the whole
track at all times preventing radon intrusion. The entire track forms a closed volume that passes through the different
layers of shielding and the veto panels. The end of the track laying outside the shielding is closed by a pneumatic gate
valve when no source is deployed, which is installed on the plate indicated in Fig.\,\ref{Fig_4}. The closed track
volume, in combination with the active purge, prevents lab air from backfilling the calibration track. Two light emitter
and detector pairs are placed on the outside of the gate valve. One pair is used to check the status of the gate valve,
while the other pair is used to verify that the radioactive line source is fully retracted behind the valve in its
storage position before closing the valve. An improper closing of the valve could damage the thin-walled line source and
in the worst case lead to a contamination of radioactive material.\\

\begin{figure}[t]
 \centering
 \includegraphics[width=0.9\columnwidth,bb=0 0 1008 756,keepaspectratio=true]{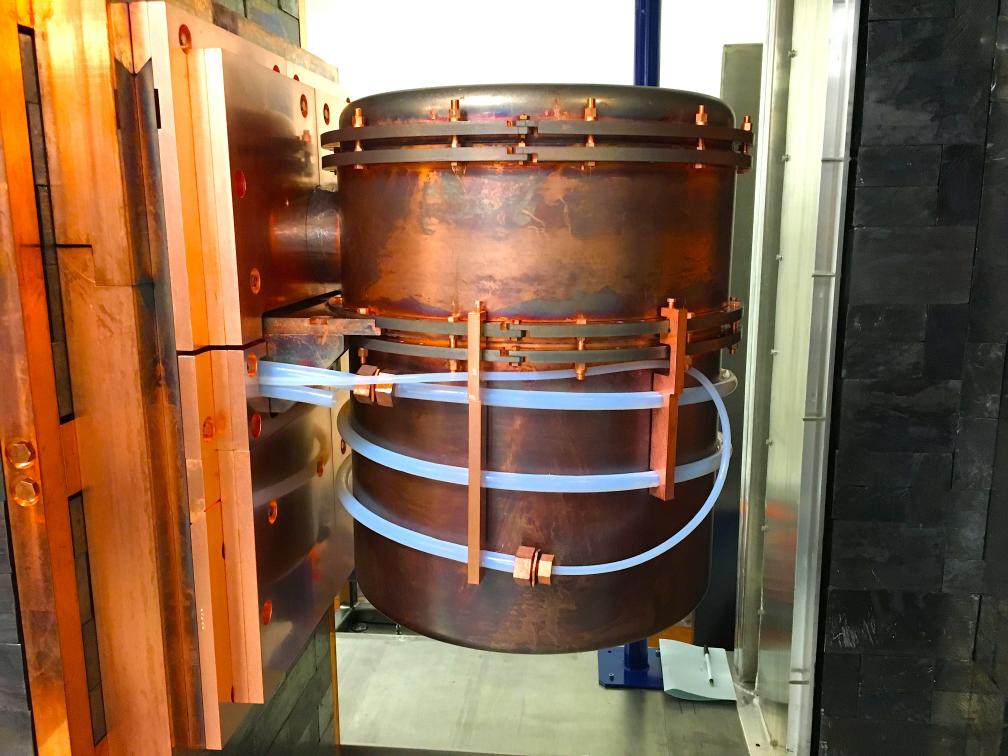}
 \caption{Inner calibration track around Module 1. The 1/4-inch-purge line is attached to the 1/2-inch calibration
track.}
 \label{Fig_3}
\end{figure}

\subsection{Source positioning}
\begin{figure}[t]
 \centering
 \includegraphics[width=0.9\columnwidth,bb=0 0 789 494,keepaspectratio=true]{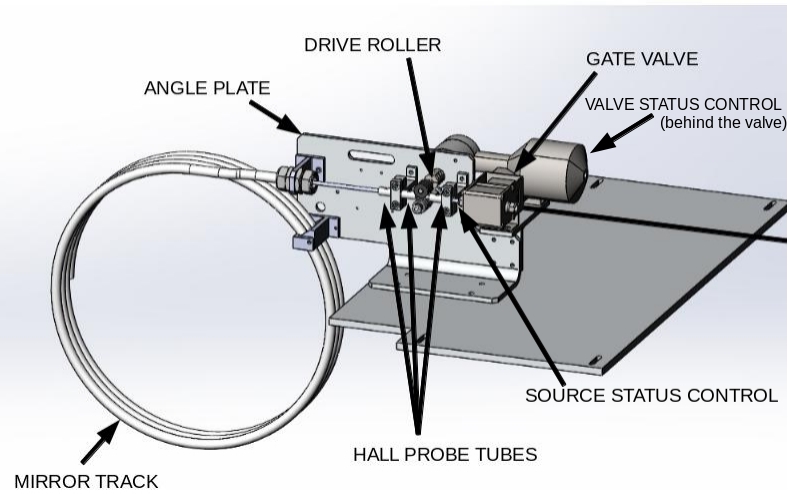}
 \caption{Schematic overview of the sensors installed at the angle plate of the calibration system. The
inner calibration track would be situated on the right side of the image.}
 \label{Fig_4}
\end{figure}

\begin{figure}[t]
 \centering
 \includegraphics[width=0.9\columnwidth,bb=0 0 1580 926,keepaspectratio=true]{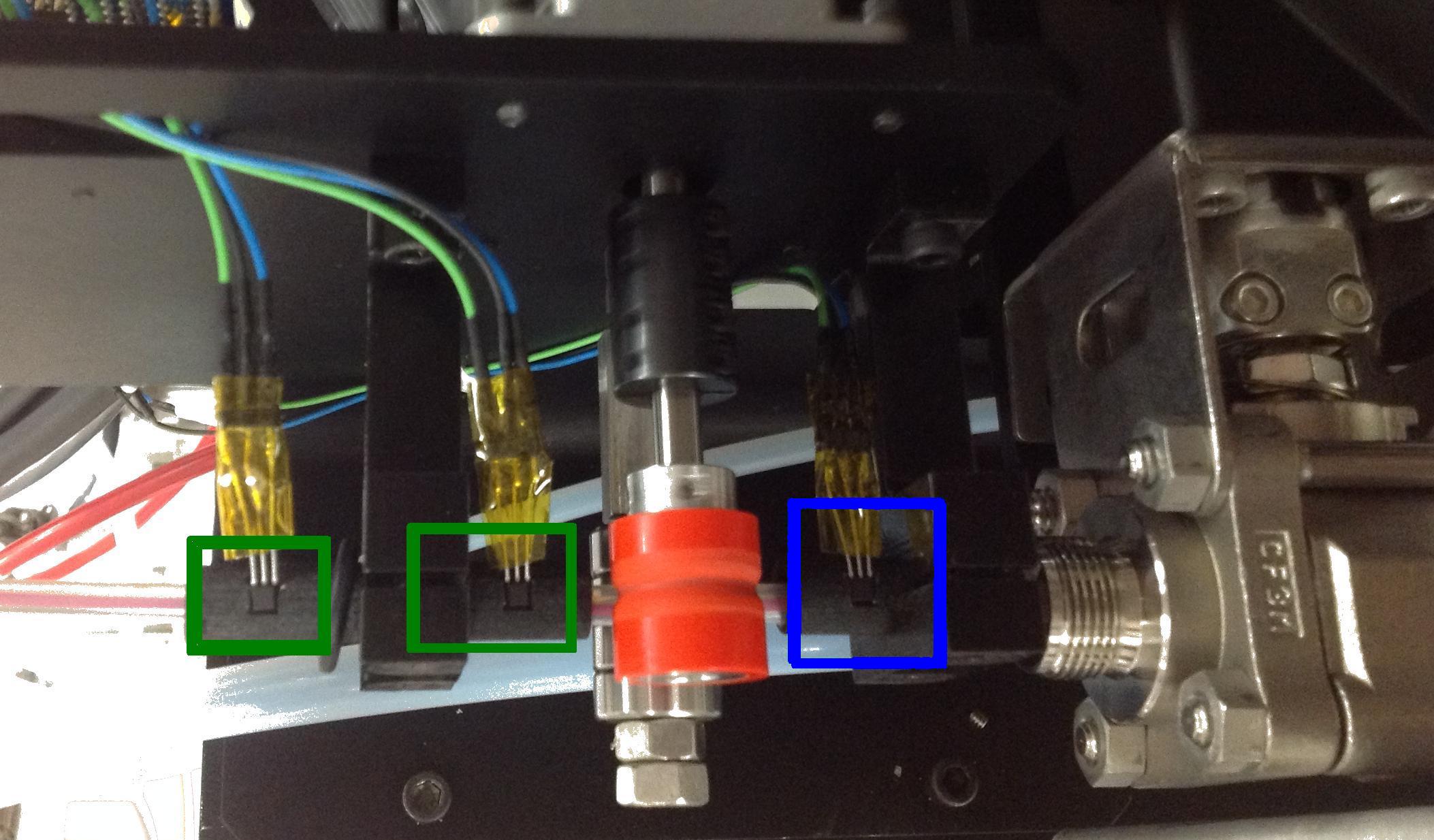}
 \includegraphics[width=0.9\columnwidth,bb=0 0 1713 1290,keepaspectratio=true]{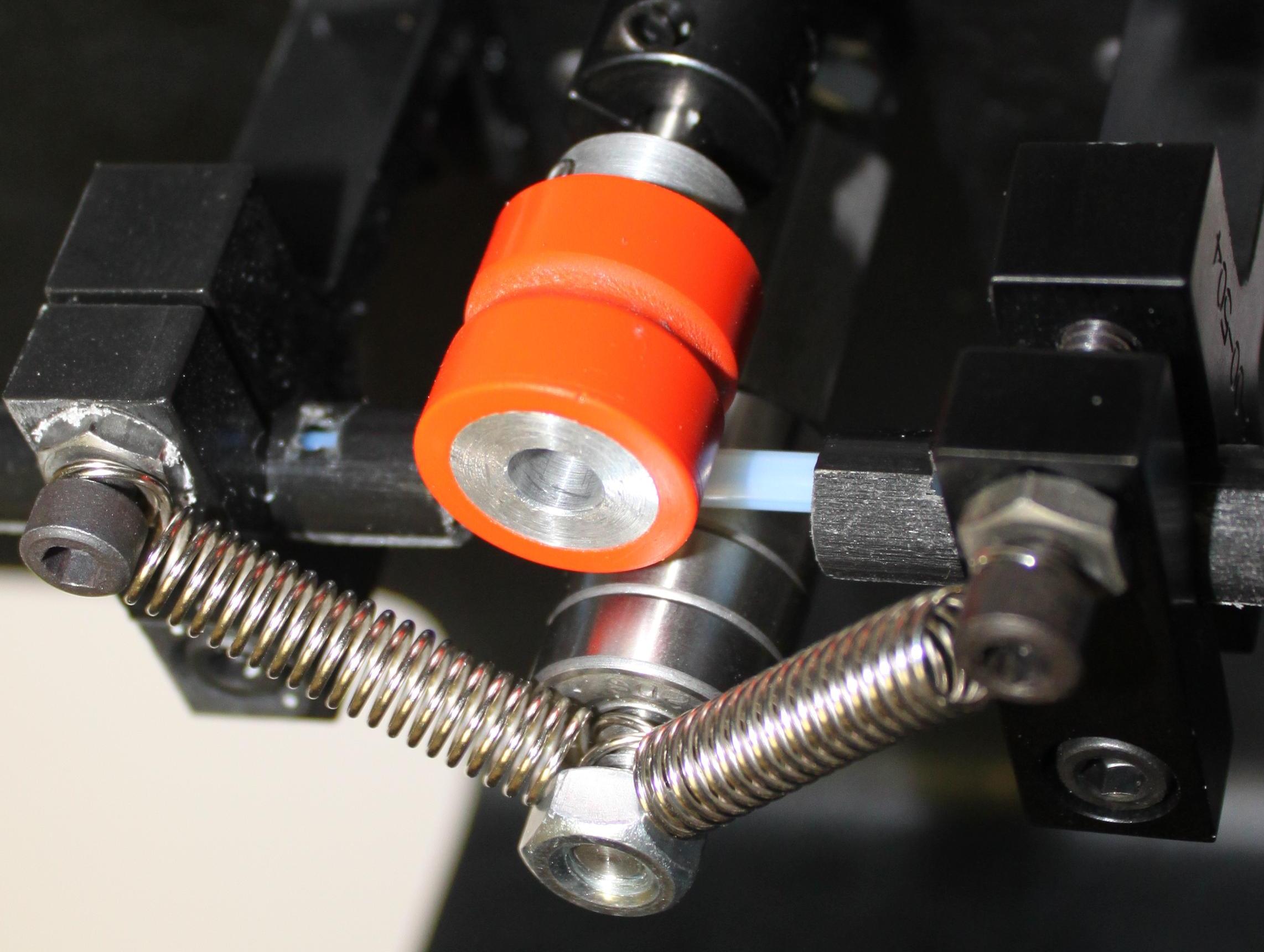}
 \caption{Radioactive line source between the two drive rollers. Two magnet sensors are placed in front of the
drive roller (green boxes, left side in the upper figure). One magnet sensor is located between drive roller and the
gate valve on the right (blue box). In the lower figure can be seen the springs that pull the idler pulley against the
drive pulley above. In both figures the shielding and the cryostat are on the right side of the figures.}
 \label{Fig_5}
\end{figure}

Inside the inner copper shielding, the  track is wrapped in a helical shape around the cryostat. The tubing is held in
place by clean electro-formed copper brackets which are attached to the cryostat.  Figure\,\ref{Fig_3} shows the
as-built configuration of the track around one module before it is moved into the shielding.\\
The source itself is guided through small tubing between two drive rollers, cf Fig.\,\ref{Fig_5}. One of
the drive rollers is rotated by a motor. The second drive roller passively rotates on an axle below the first one.
The axle is pulled by two springs against the actively rotating drive roller so that the line source is
gently clamped in-between the two drive rollers. This ensures a long-term grip of the drive rollers on the line
source.\\
A deployed source wraps twice around the cryostat allowing a simultaneous calibration of all detectors, in contrast
to for example the calibration system of \textsc{Gerda} \cite{Baudis2013}. Simulations have shown that individual
detector rates are highly sensitive to a source's position within the track. More specifically, detector rates are
heavily dependent on how
far into the helical track a source is deployed. To reproduce the source position more reliably, we have developed a
positioning system using Hall effect switches\footnote{Infineon TLE4905L E6433}. Three magnetic
sensors, see Fig.\,\ref{Fig_4} , are connected to an Arduino UNO
\footnote{https://www.arduino.cc/en/Main/Products}. These sensors register a passing magnet in a moving line source
via the Hall effect. By reading their signal continuously, the arrangement of three sensors outside and two magnets
inside the source allows us to determine the status of the source and reproduce its position when deployed within a few
millimeters. A special sequence in the readout of the magnetic sensors gives feedback to the experiment control if the
source is deployed, in deployment, retracted or in retraction.\\
Sources are stored in mirror tracks behind the last magnet sensor at the outer end of the calibration system plate.
This mirror track, made of the same PTFE tubing as the other parts of the track, is wound similarly to the track inside
the shield, thereby ``mirroring" the shape of the deployed source as it
is stored. It
is possible to access this part of the calibration system and to exchange the whole mirror track with the source inside
if another calibration source needs to be used. In the final configuration the $^{228}$Th is installed as standard
source. It can be exchanged with the $^{60}$Co line source if additional commissioning is needed.

\subsection{System controls}
The sensors attached to the Arduino controller and the motor can be controlled individually via the data acquisition
(DAQ) software \textsc{ORCA}\,\cite{How04} in an expert mode which allows debugging and tests. Non-expert users can use
an interface with reduced functionality on the DAQ machines.
The latter approach can start a calibration of the modules with a mouse click with automated control of calibration
run duration, run-bit settings indicating a calibration run and supervision of the DAQ during deployment, calibration,
and retraction of the sources. \\

\subsection{Interference and background contribution}
The copper cross-arm of the cryostat delivers the pumping and cooling power to the cryostat, but therefore also provides
a direct line-of-sight through the lead shield to the detectors. The track was designed so that the source would not lie
in this line-of-sight when place in its storage position. While going through the outer layers of lead,
the track is curved. The mirror track itself is coiled vertically so that the source in storage is
located below the cross-arm. Tests with button sources at certain locations behind the shield have verified that a
source in storage does not contribute to the background of the \textsc{Majorana Demonstrator}.\\
We considered the possibility of removable contamination from sources being carried into or left inside the tube next to
the cryostat. The mirror tracks and the cover of the system create an extra shielding so that direct contact to the
sources due to ongoing lab work is avoided and line sources stay clean. In addition, the line sources, part of the
tracks, and drive rollers are wiped on a regular basis and have not shown abrasion or any signs of dirt. During the last
year of running we have not found any increase of backgrounds attributable to the calibration source.

\section{Commissioning and performance}
\begin{figure}[t]
 \centering
 \includegraphics[width=1.0\columnwidth,keepaspectratio=true]{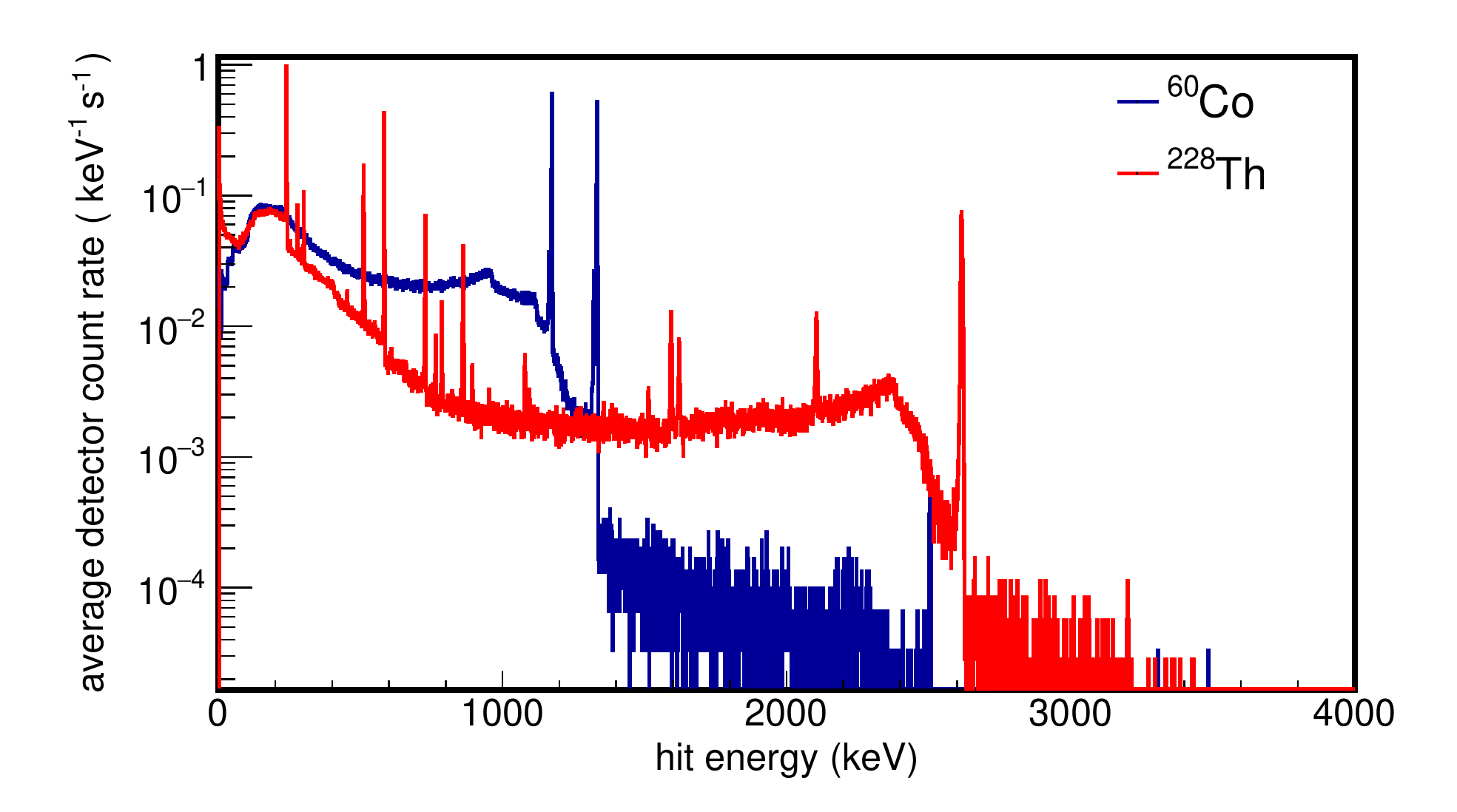}
 \caption{Single-detector energy spectrum of Module 1, averaged over the number of
detectors and with a calibration time of an one hour-long data set. The spectra are shown for the detectors biased
at the time of the measurement.}
 \label{Fig_6}
\end{figure}

\begin{figure}[t]
 \centering
 \includegraphics[width=1.0\columnwidth,keepaspectratio=true]{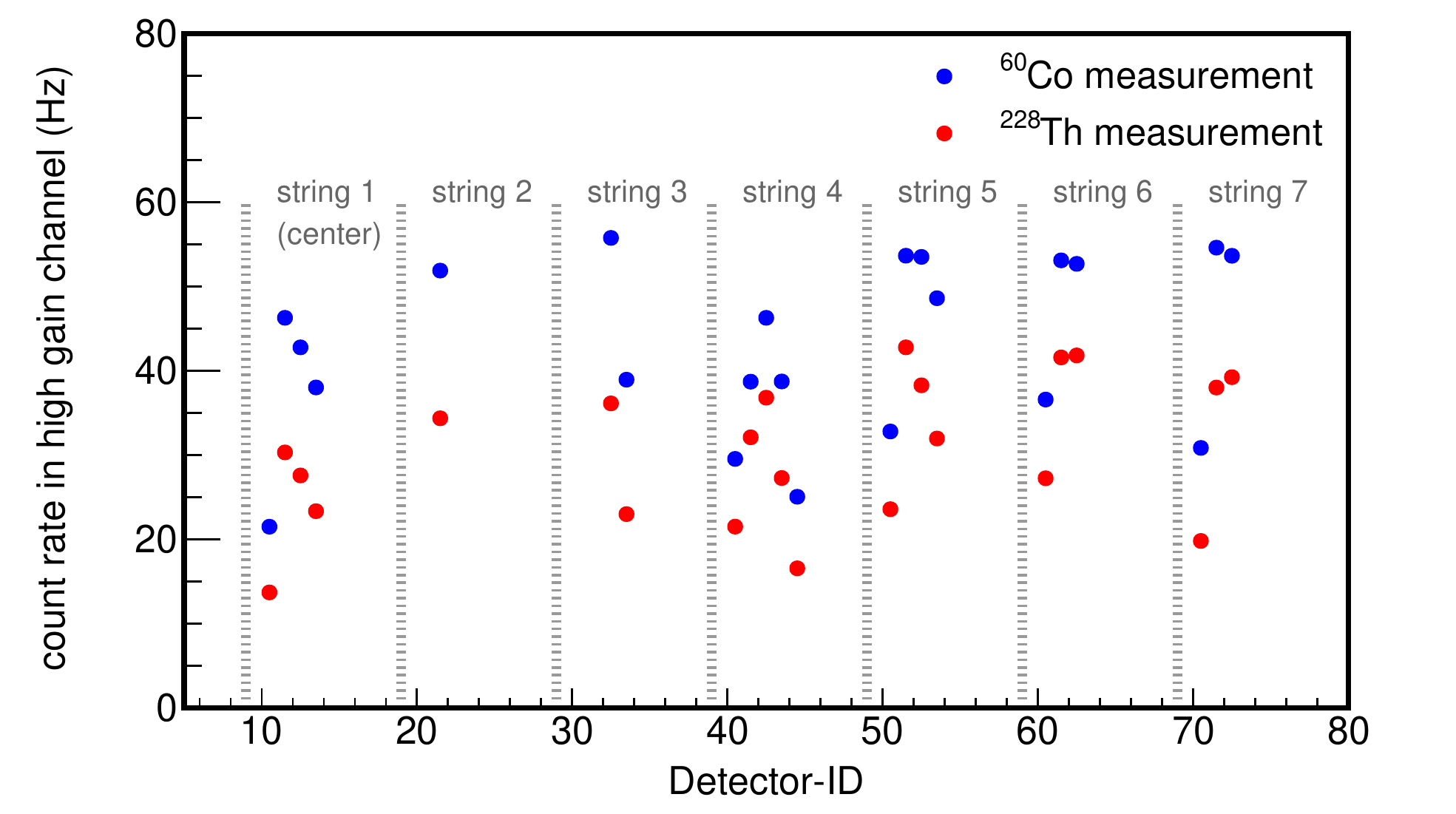}
 \caption{Integral count rate per detector in the high gain channel between 20 and 4000 keV for the two installed line
sources. The count rate is given for the detectors used for the first data set of Module 1. }
 \label{Fig_7}
\end{figure}

\begin{figure}[t]
 \centering
 \includegraphics[width=1.0\columnwidth,keepaspectratio=true]{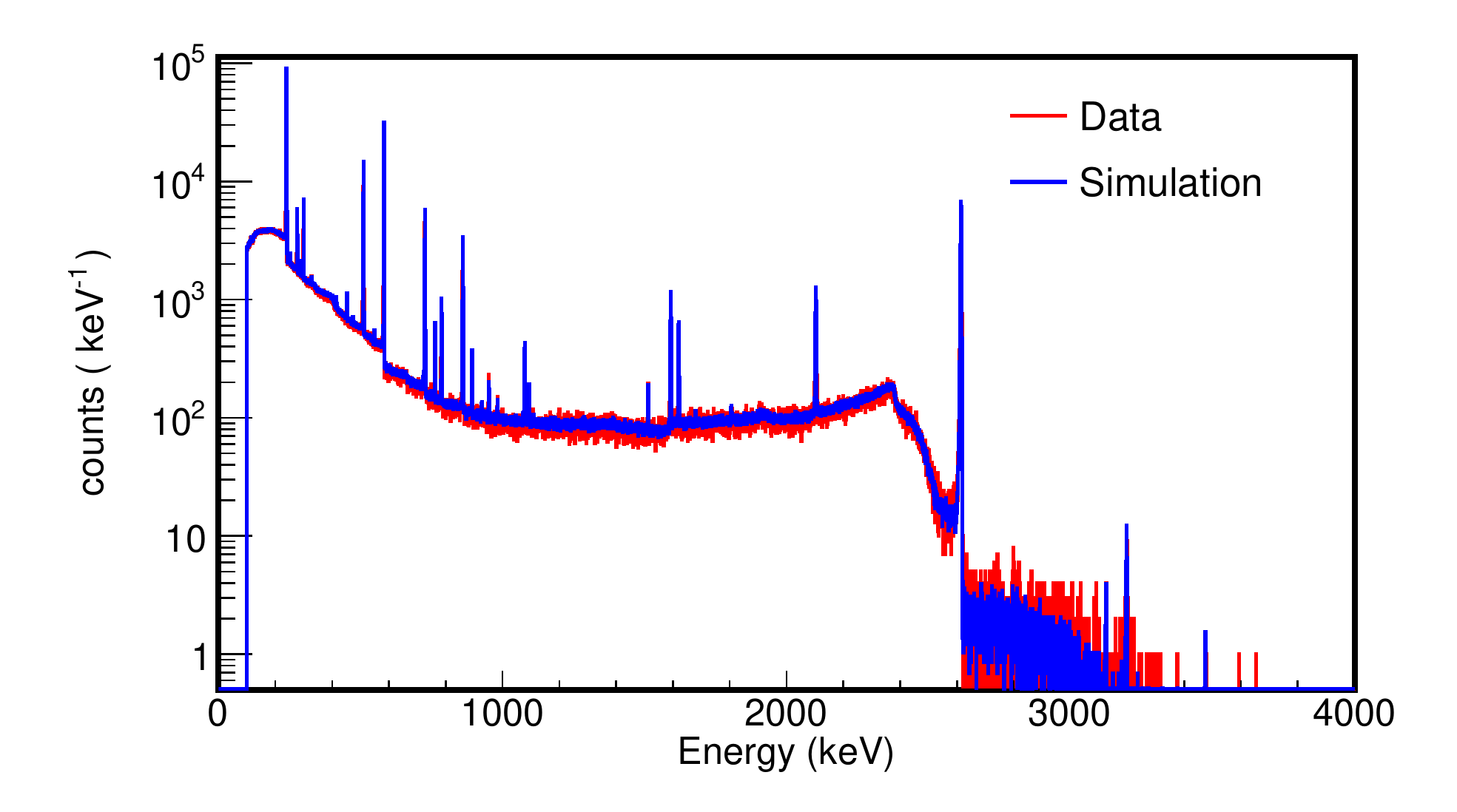}
 \caption{Comparison of a $^{228}$Th line source simulation using \textsc{MaGe} and a measurement of Module 1. The
simulated distribution was normalized by matching the integrals of both curves in the range from 2595\,keV to
2635\,keV.}
 \label{Fig_8}
\end{figure}

During commissioning, a stress test of the system was performed. Calibration sources were deployed and retracted
about a hundred times without any incident. Assuming a weekly calibration over the expected 5-year lifetime of the
\textsc{Majorana Demonstrator} the number of operational cycles is about in the same order.  The software was tested and
different failure or mishandling modes were investigated. An operational procedure was written. \\
Figure\,\ref{Fig_6} is taken from a data set from 2015 in which the first module was situated inside the
shielding. In this data set, detectors in Module 1 were calibrated using its line sources. A spectrum
with several peaks at different energies is available for calibration. Fig.\,\ref{Fig_7} shows the integral count
rate for energies between 20 and 4000 keV. The average count rate for calibrations with the $^{60}$Co and the $^{228}$Th
source is around 40\,Hz and 30\,Hz, respectively. This fulfils the requirement on the maximum count rates discussed
before.  The figure shows that several detectors closer to the line source path have higher
rates, as expected. Detectors at the inside or further away from the path have slightly lower rates. However, the built
geometry reached the goal to have a balanced count rate within the array. \\
Of course, a deployed source around one module can be seen by detectors in the second cryostat. The recorded spectra in
these detectors can be used to test stability. However, the count rate is not sufficient for a full calibration. For a
full calibration data set of one module, a dedicated deployment of the corresponding source is necessary. In addition,
it is possible to deploy both sources at the same time for high-rate data tests. \\
Calibration measurements are also used as a reference for our simulation campaign. The simulations using the
Geant4-based framework \textsc{MaGe}\,\cite{Bauer2006, Boswell2010} are compared to measurements in order to
validate the implementation of the geometry. Various parameters like the height-ratio between the full-energy
peaks, the full-energy peak to Compton continuum ratio and multiplicity distributions are affected by the positioning of
the source relative to the detectors as well as by the amount of material located between source and detectors.
Using the calibration measurements, the transition layer of the models in the simulations can be validated. The size of
the transition layer affects the size of the step function under a full-energy peak which is necessary to get similar
spectral shape. A first analysis showed that \textsc{MaGe} is able to describe the measured spectra in
calibration runs very well, as shown in Fig.\,\ref{Fig_8}. The simulation is in very good agreement and can be used to
build up a pulse-shape library, to study the efficiency of cuts and the background in general.

\section{Summary}
A calibration system has been designed for the \textsc{Majorana Demonstrator}. The system has to fulfil
special requirements in terms of cleanliness, stable positioning and reproducibility. The strength of the radioactive
source was adjusted to the special needs of the data stream and ensures stable data taking.
Two systems, one for each cryostat in the shield, were built. The systems deploy radioactive sources from
outside the shielding to a position next to the detectors so that energy calibrations and other studies are
possible. In contrast to other systems, our line source approach can calibrate the whole array at the same time
without moving the source. The results can be used to study the agreement of coincidences in experiment
and simulation. The design, using a line source in which the radioactivity is distributed along a certain length,
ensures a sufficient count rate in all detectors of the array so that one calibration run can be used for all detectors
of a module. Also, it has been shown that the calibration system around one module can easily be copied, which is
important for the scalability of such a modular approach when going to ton-scale experiments.
Both systems are operated remotely and the controls are implemented in the DAQ system of the \textsc{Majorana
Demonstrator}. While the exact design depends on the future experimental conditions, the MJD calibration system shows
that it is possible to construct a system using easily accessible materials, sensors and readout electronics while
still holding to all cleanliness and low-background limits.

\section{Acknowledgements}

This material is based upon work supported by the U.S. Department of Energy, Office of Science, Office of Nuclear
Physics under Award  Numbers DE-AC02-05CH11231, DE-AC52-06NA25396, DE-FG02-97ER41041, DE-FG02-97ER41033,
DE-FG02-97ER41042, DE-SC0012612, DE-FG02-10ER41715, DE-SC0010254, and DE-FG02-97ER41020. We acknowledge support from the
Particle Astrophysics Program and Nuclear Physics Program of the National Science Foundation through grant numbers
PHY-0919270, PHY-1003940, 0855314, PHY-1202950, MRI 0923142 and 1003399. We acknowledge support from the Russian
Foundation for Basic Research, grant No. 15-02-02919. We  acknowledge the support of the U.S. Department of Energy
through the LANL/LDRD Program. This research used resources of the Oak Ridge Leadership Computing Facility, which is a
DOE Office of Science User Facility supported under Contract DE-AC05-00OR22725. This research used resources of the
National Energy Research Scientific Computing Center, a DOE Office of Science User Facility supported under Contract No.
DE-AC02-05CH11231. We thank our hosts and colleagues at the Sanford Underground Research Facility for their support.

\bibliography{MJDCalibrationSystemV8}

\end{document}